# Multi-refractive-index metamaterials based on subwavelength waveguide arrays

Zhaoning Yu[1,3], Chenghao Wan[2], Jad Salman[3], Bradley S. Gundlach[3],
Yuzhe Xiao[3], Zongfu Yu[3], Mikhail A. Kats[3,2,1*]

[1]Department of Physics, [2]Department of Materials Science and Engineering, [3]Department of Electrical and Computer Engineering, University of Wisconsin-Madison, Madison, Wisconsin, USA 53706

*Email address: mkats@wisc.edu

**We demonstrate a metamaterial that cannot be described by a single set of refractive-index and impedance values, even for fixed frequency and polarization. The metamaterial structure is a stack of dissimilar waveguides with subwavelength width and spacing, which guide light at different phase velocities. This multi-refractive-index metamaterial (MRIM) can be viewed as a spatial superposition of multiple homogeneous materials. Using full-wave simulations, we demonstrate several optical components based on MRIMs, including prisms that deflect light to multiple angles, lenses with multiple focal points, and multi-index Fabry-Perot etalons with an enhanced density of resonant modes. We provide a theoretical framework for analyzing MRIMs to determine effective refractive indices, fractions of power to each channel (*i.e.*, to each refractive index), and transmittance.**

In conventional refractive optics, a complex refractive index can typically be assigned to the various constituent materials to describe light-matter interaction. The refractive index of materials is frequency-dependent (dispersion), and can have a polarization dependence (birefringence and dichroism). However, for a given frequency and polarization along some axis of symmetry, a conventional material has only one well-defined value of the refractive index.

The study of metamaterials—artificial materials comprising subwavelength components—has resulted in the demonstration of many optical properties that are not found in nature [1][2][3]. Similar to conventional materials, a metamaterial is typically treated as a homogeneous material with a well-defined effective refractive index (and impedance), following an effective medium theory [4]. Both theoretical [5] and experimental [6] research has been carried out to determine the effective index of different metamaterials, which is generally a single complex value given a particular frequency and linear polarization state along an axis of symmetry.

Here, we demonstrate optical components based on a structure that cannot be ascribed a single refractive index value for a given frequency and polarization, even though it comprises an array of elements that are deeply subwavelength. By carefully packing well-separated subwavelength optical channels, the metamaterial can be made to have multiple simultaneous refractive indices that can be engineered separately. The extra degree of freedom of the refractive index (beyond dispersion and birefringence) leads to the generalization of Snell's law and related well-known optical phenomena, modifying the behavior of common refractive components such as prisms and lenses comprised of such a metamaterial.

Consider light refracted by a prism with incident angle $\theta_i$, as shown in Fig. 1. Snell's law relates the refracted angle ($\theta_t$) of light in medium $t$ with the incident angle ($\theta_i$) in medium $i$: $n_i sin\theta_i = n_t sin\theta_t$, where $n_i$ and $n_t$ are the scalar refractive indices in $i$ and $t$, respectively. Here, we consider a prism made from a hypothetical multi-refractive-index metamaterial (MRIM), such that $n_i$ is transformed into a vector containing multiple scalar effective refractive indices ($n_{eff,1}, n_{eff,2}, ...$)—at a single frequency and polarization—resulting in multiple refracted angles ($\theta_{t,1}, \theta_{t,2}, ...$). Thus, a more general form of Snell's law to describe refraction from a MRIM to free-space is:

$$\begin{bmatrix} n_{eff,1} \\ n_{eff,2} \\ ... \end{bmatrix} sin\theta_i = n_t sin \begin{bmatrix} \theta_{t,1} \\ \theta_{t,2} \\ ... \end{bmatrix} \tag{1}$$

For example, we assume a prism comprising a MRIM with period $D$ and effective indices of $n_{eff,1} = 1.5$ and $n_{eff,2} = 2.5$ (Fig. 1(b)). When p-polarized long-wavelength ($\lambda \gg D$) light of incident angle $\theta_i = 20°$ is refracted from the prism to free space, the wave is split into two, corresponding to two refracted angles given by Eq. 1 (*i.e.,* $\theta_{t,1} = 31°$ and $\theta_{t,2} = 59°$).

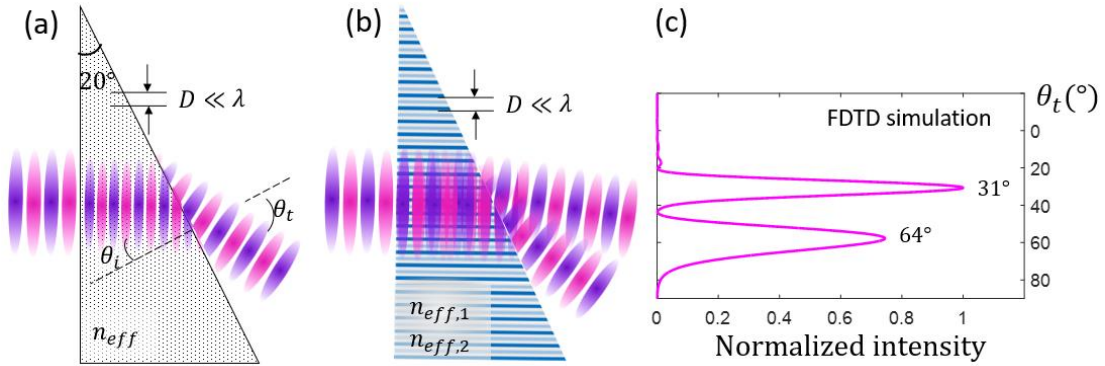

Fig 1. Linearly polarized monochromatic light with incident angle $\theta_i$ is refracted by a prism/air interface, where the prism comprises (a) a conventional material or metamaterial, resulting in a single refracted angle, $\theta_t$, or (b) a multi-refractive-index metamaterial (MRIM) with period $D$, generating two distinct refracted beams, corresponding to two peaks in (c) the far-field angular distribution of the refracted light. (c) is simulated using the finite-difference time-domain (FDTD) method with incident light being a transverse-magnetic (TM) Gaussian beam with free-space wavelength $\lambda_0 = 8 \: \mu m$, and with the prism being a subwavelength metal-insulator-metal (MIM) waveguide array with parameters described in the text.

To achieve multiple simultaneous refractive indices in a MRIM, light must propagate with multiple phase velocities. We accomplish this by designing a structure with multiple well-separated subwavelength propagation modes with different propagation constants. Fig. 2(a) shows our MRIM comprising alternating subwavelength metal-insulator-metal (MIM) waveguides [7][8] with different mode indices and thus different phase velocities. In each subwavelength waveguide, only one mode is allowed to propagate. Because of the simultaneous presence of multiple subwavelength channels with differing phase velocities, the resulting periodic structure must be described by multiple refractive indices at the same time, instead of a single effective index based on standard effective medium theory [5]. The dispersion of this class of structures was previously studied using the transfer-matrix method by Orlov et al., who looked for complex topologies in the dispersion surfaces [9]. Conversely, we focus on the simplest dispersion features of such structures, and investigate how they can be used to realize new types of refractive and interference optics.

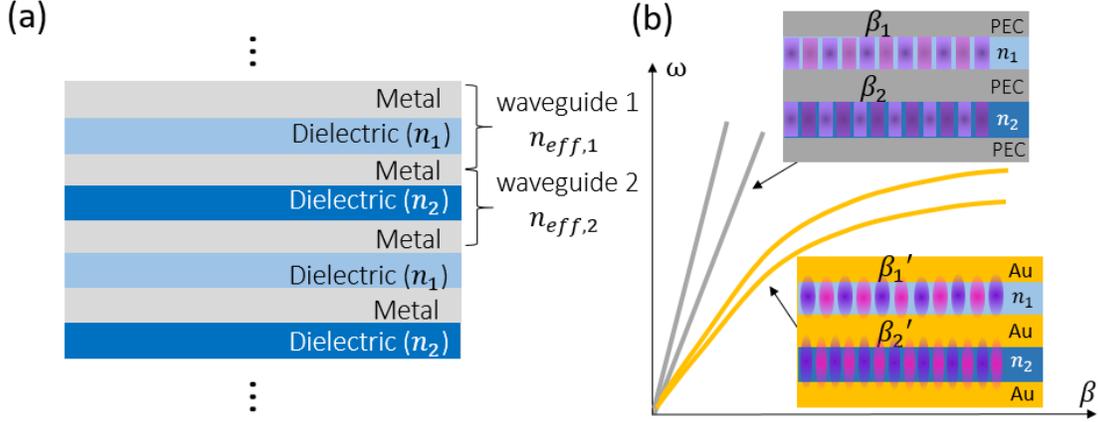

Fig 2. (a) A periodic metamaterial comprising multiple deep-subwavelength waveguides with different mode indices, resulting in multiple simultaneous values of refractive index (here, two values). The metal-insulator-metal (MIM) waveguides comprise two metal layers and one dielectric layer in between with refractive index $n_1$ or $n_2$. (b) Sketch of the double-branch dispersion curve (with angular frequency $\omega$ and propagation constant $\beta$) of such a MRIM, with the metal being a perfect electric conductor (PEC) or gold (Au). For PEC walls, the allowed mode in each MIM waveguide is the transverse electromagnetic (TEM) mode, whereas for gold walls, this mode is the antisymmetric coupled surface plasmon polariton (SPP).

To demonstrate the multiple refraction shown schematically in Fig. 1(b), we consider a MRIM prism comprising two types of subwavelength MIM waveguides, with 100-nm walls made from a perfect electric conductor (PEC) and two types of dielectric layers: the first with $n_1 = 1.5$ and a thickness of 100 nm, and the second with $n_2 = 2.5$ and a thickness of 150 nm. The TEM mode of an MIM waveguide with PEC walls has no cutoff frequency and a linear dispersion curve (Fig. 2(b)), where the mode index is equal to the refractive index of the dielectric layer ($n_{eff,1} = n_1 = 1.5$, $n_{eff,2} = n_2 = 2.5$). The combination of thickness and index of the dielectric was chosen such that light tended to be coupled equally into the different channels, as described below. We initially selected lossless PEC for the walls so that we could investigate MRIM behavior without attenuation. However, since there is no loss, additional care is needed in any simulation to prevent reflected light from bouncing in the prism and forming secondary refracted beams, complicating the analysis.

We performed finite-difference time-domain (FDTD) simulations, which showed that the far-field distribution of the refracted light has peaks at $\theta'_{t,1} = 31°$ and $\theta'_{t,2} = 64°$ (Fig. 1(c)), compared to $31°$ and $59°$ from the ray-optics-like calculation in Eq. 1. The 8% shift in $\theta_{t,2}$ drops to $< 2\%$ if the wavelength is doubled (simulated far field distribution shown in the supplementary material).

Because the dispersion diagram of the MRIM contains several curves, we can view the MRIM as a spatial superposition of several distinct refractive media, each with a single dispersion curve. In particular, the MRIM in Fig. 2 can be regarded as the superposition of two hyperbolic metamaterials [1]. As a demonstration of this superposition effect, we simulated a cylindrical lens comprising a MRIM with $n_{eff,1} = 2.1$ and $n_{eff,2} = 4.5$ (Fig. 3(a, c)), and observed the focusing of light to two focal points. The field distribution of light focused by this lens is almost identical to the coherent sum of the fields of two conventional cylindrical lenses of the same size and shape, comprising uniform dielectrics with these two refractive indices (Fig. 3(b)).

The MRIM lens in Fig. 3(a, c) was designed with gold walls, using the optical properties for gold from ref. [10]. As a result, the mode indices of the MIM waveguides and hence the effective indices of the MRIM are larger than the indices of the dielectric layers themselves ($n_{eff,1} > n_1 = 1.5$, $n_{eff,2} > n_2 = 4$) [8]. Gold was selected to more-closely mimic experimentally realizable conditions, and also because we found that truncated PEC waveguides resulted in Fabry-Perot-like resonances (described in more detail in Fig. 4(b)), negatively affecting focusing performance. The combination of index and thickness of the dielectric layers was chosen so that the two foci of the MRIM lens are separable and of comparable field magnitude (procedure details in the supplementary material).

The focal lengths of a MRIM lens can be approximately predicted using the Lensmaker's Formula [11], generalized for multi-index materials:

$$\begin{bmatrix} f_1^{cal} \\ f_2^{cal} \\ \ldots \end{bmatrix} = \frac{R}{\begin{bmatrix} n_{eff,1} \\ n_{eff,2} \\ \ldots \end{bmatrix} - 1} \tag{2}$$

where $R$ is the radius of curvature. We note that for small lenses (with Fresnel number $N < 10$), like the one in our simulation (Fig. 3), the actual focal length will be smaller than the prediction of Eq. 2 due to diffraction effects [12]. After correcting for diffraction, we obtained calculated focal lengths of 24 μm and 67 μm (see supplementary), compared to the simulated focal lengths of $f_1 = 27$ μm and $f_2 = 59$ μm (Fig. 3(a)). The main difference between theory and simulation is that each theoretical focal length is calculated separately, while in the simulation the MRIM lens generates one overlapping field which shifts the two foci closer to each other when the fields are added coherently. After summing the fields of focused light from two dielectric lenses with the same size, shape, and refractive indices as the MRIM lens in Fig. 3(a), the focal lengths of the "superimposed lens" are found to be $f_1' = 29$ μm and $f_2' = 60$ μm (Fig. 3(b)), which gives a better prediction of the foci of the MRIM lens.

In order to test the MRIM lens for imaging purposes, we also simulated the MRIM lens with light incident at an angle of 10° (Fig. 3(c)), where two off-axis foci are generated at $f_1 = 36$ μm and $f_2 = 57$ μm, compared to the coherently summed fields of the two superimposed dielectric lenses which result in $f_1' = 39$ μm and $f_2' = 63$ μm (the latter shown in the supplementary material). The 10% difference between the focal length of the MRIM lens and that of the "superimposed lens" may be a result of the different transmission efficiencies of the two channels in the MRIM, which do not correspond to the relative transmission efficiencies of the two dielectric lenses.

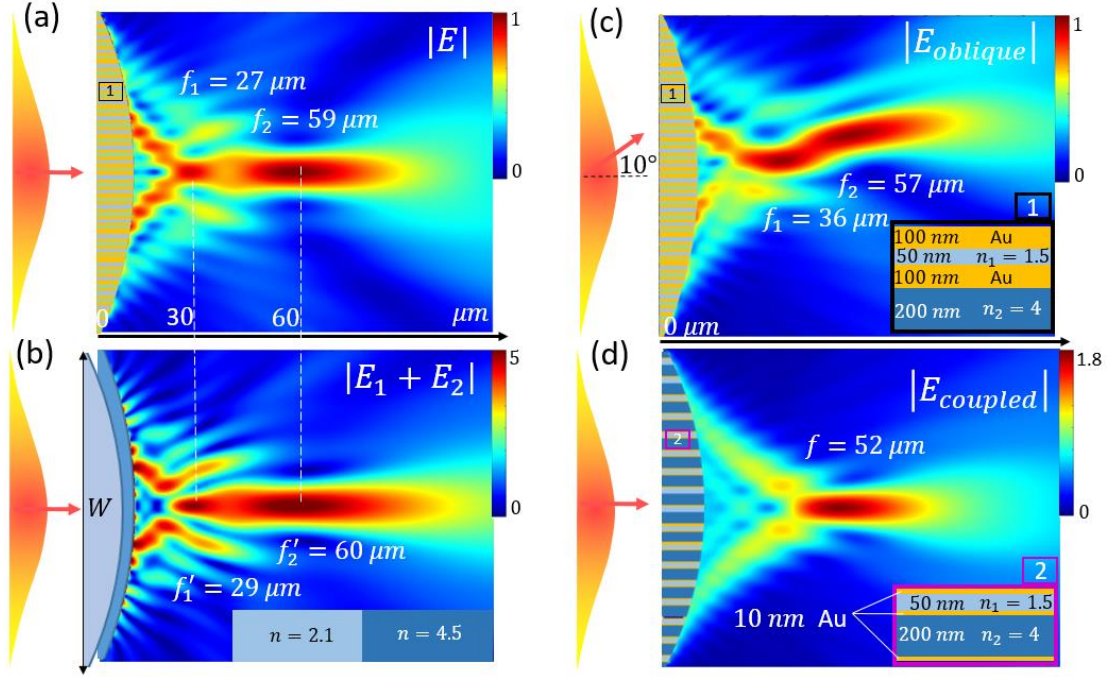

Fig 3. FDTD simulations of light focused by cylindrical lenses. (a) & (c): Two foci are generated by our lens comprising a MRIM with optically thick gold walls, shown in the inset of (c), for (a) normal and (c) oblique incidence at an angle of 10° from free space. (b): A field distribution similar to that in (a) is found by coherently adding the fields from two lenses of the same size and shape as in (a, c), comprising two different homogeneous transparent dielectrics with refractive indices corresponding to the two effective indices of the MRIM. (d): After reducing the gold-layer thickness, only one focus can be observed. All lenses are of the same plano-convex shape, with radius of curvature $R = 85\ \mu m$, width $W = 80\ \mu m$, and minimum thickness $t_e = 1\ \mu m$ at the edge. The incident light is a p-polarized Gaussian beam at $\lambda_0 = 8\ \mu m$, with beam waist $w_0 = 35\ \mu m$. The refractive index of gold is taken to be $n_{Au} = 8.5 + 46.4i$ [10].

To analyze the efficiency of MRIM-based optical components, such as the lens in Fig. 3, we need to calculate the transmission coefficient at the interface between free space and a MRIM, where standard Fresnel equations do not apply due to the presence of multiple refractive indices and the nontrivial wave impedance of the MRIM. Thus, we derive modified Fresnel equations at interfaces involving a MRIM by considering the boundary conditions and conservation of power. When light (with original electric field $E_0$) propagates from free space to a MRIM, it can be reflected (with field $E_r$) or enter the optical channels of the MRIM (with $E_j$ representing the field inside the $j^{th}$ channel), as drawn in Fig. S3(a) of the supplementary material. For s-polarized light, $E_j$ is equal to the electric field in the metal walls at the dielectric-metal boundary inside the MRIM, and thus is negligible, meaning s-polarized light can barely enter the MRIM (i.e., the reflectance $R_s \approx 1$). For p-polarized light, the magnetic field $H$ is continuous from free space to each optical channel. For simplicity, here we only consider a MRIM comprising PEC walls, where the magnetic field is uniform across each waveguide (i.e., only the TEM mode can propagate) and linearly related to the product of the electric field and the refractive index of the dielectric ($H_j = E_j n_j / Z_0$, $Z_0$ is free space wave impedance). The magnetic field in free space, which includes the incident field and the reflected field (with a $\pi$ phase shift), is also linearly related to the product of the electric field and the refractive index ($n_{free\ space} = 1$) by the same factor, *i.e.*, $H_0 - H_r = (E_0 - E_r)/Z_0$. Subsequently, the continuity of the magnetic field ($H_0 - H_r = H_1 = H_2 = \cdots = H_j$) can be written as:

$$E_0 - E_r = E_1 n_1 = \cdots = E_j n_j = \cdots \tag{3}$$

Note that for a MRIM using real metals, Eq. 3 must be generalized for non-uniform mode profiles. Simultaneously with Eq. 3, power must be conserved at the interface, such that the incident power is equal to the sum of the reflected power plus the transmitted power into all of the channels ($P_{inc} = P_r + \sum_j P_j$). Replacing the magnetic field ($H$) with the electric field ($H \propto En$), we can write the power flow as $P = EH(\text{Area}) \propto nE^2(d\cos\theta)$. Therefore, we obtain:

$$\cos\theta_i DE_0^2 = \cos\theta_i DE_r^2 + \sum_j n_j d_j E_j^2 \tag{4}$$

where $\theta_i$ is the angle of incidence in free space and $d_j$ the width of the $j$th type of waveguide (Fig. S3(a) of the supplementary material). Combining Eqs. 3 and 4, we obtain the Fresnel-like transmission and reflection coefficients for p-polarized light propagating from free space (denoted in subscript as "$f$") to a PEC-based MRIM:

$$t_{fj} \triangleq \frac{E_j}{E_0} = \frac{2\cos\theta_i D/n_j}{\cos\theta_i D + \sum_k d_k/n_k} \tag{5-1}$$

$$r_{ff} \triangleq \frac{E_r}{E_0} = -\frac{\cos\theta_i D - \sum_k d_k/n_k}{\cos\theta_i D + \sum_k d_k/n_k} \tag{5-2}$$

where $t_{fj}$ is the field transmission coefficient into the $j$th channel. We have verified Eq. 5 by FDTD simulations with different angles of incidence $\theta_i$, thicknesses ($d_j$), and indices ($n_j$) of optical channels (see the supplementary material Fig. S3).

Though Eq. 5 is only directly applicable for PEC walls, we can still use it to estimate the reflection at the left boundary of the MRIM lens with gold walls in Fig. 3(a), which gives ~40% reflection. Following absorption in the gold during propagation through the lens, the simulation shows only ~5% of the total incident power is transmitted and focused. We note that there are many possible routes to decreasing the absorption losses in such a structure (e.g., ref. [13]), though we do not explore them here. One feasible way to increase the overall efficiency is to decrease the thickness of the gold walls so that the MRIM has impedance closer to that of free space, which lowers the reflection. However, the gold walls cannot be too thin, otherwise the modes will couple and a single effective index can then be assigned to the metamaterial [14]. This is shown in Fig. 3(d), where the gold layer thickness is reduced significantly compared to Fig. 3(a); the structure can then be assigned a single effective index $n_{eff} = 2.3$, and only one focus is observed, which lies in between the two foci from Fig. 3(a).

We further calculate the transmission and reflection coefficients when light is incident from within a MRIM onto an interface with free space. This problem is more complex than the case of incidence from free space, because in addition to reflection or transmission, "swapping" can occur, *i.e.*, light can jump from one optical channel ($j$) to another ($l \neq j$), as shown in the supplementary Fig. S4(a). This can be seen from the continuity of magnetic fields at the interface between the MRIM and free space. Similar to Eq. 3, we replace the magnetic field with the product of the refractive index and the electric field:

$$E_0 n_j = E_t = E_1 n_1 \ldots = E_l n_l = \cdots \tag{6}$$

The TEM mode in channel $j$ is transmitted to free space (with electric field $E_t$), which also induces a nonzero electric field at the $l^{th}$ channel ($E_l = E_t/n_l \neq 0$). As a result, part of the incident power is swapped:

$$d_j n_j E_0^2 = d_j n_j E_r^2 + D E_t^2 + \sum_{l \neq j} n_l d_l E_l^2 \tag{7}$$

With Eqs. 6 and 7, we obtain the Fresnel-like transmission, reflection, and swapping coefficients for light propagating from the $j^{th}$ optical channel of a MRIM to free space or the $l^{th}$ optical channel:

$$t_{jf} \triangleq \frac{E_t}{E_0} = \frac{2 d_j}{\sum_k d_k/n_{eff,k} + D} \tag{8-1}$$

$$r_{jj} \triangleq \frac{E_r}{E_0} = \frac{\sum_{k \neq j} d_k/n_{eff,k} + D - d_j/n_{eff,j}}{\sum_k d_k/n_{eff,k} + D} \tag{8-2}$$

$$s_{jl} \triangleq \frac{E_l}{E_0} = \frac{-2 d_j/n_{eff,l}}{\sum_k d_k/n_{eff,k} + D} \tag{8-3}$$

A comparison between Eq. 8 and full-wave simulations can be found in the supplementary material Fig. S4.

To further examine the Fresnel-like equations derived for the interface between free space and a MRIM, we simulated a Fabry-Perot-like etalon comprising a lossless MRIM with two types of optical channels (Fig. 4(a)). Taking results from Eqs. 5 and 8, we can calculate the total transmission coefficient through the two-channel etalon (detailed calculation and coefficients of etalon with more than two channels in the supplementary material):

$$\frac{E_t}{E_0} = \begin{bmatrix} t_{f1} e^{i\phi_1} & t_{f2} e^{i\phi_2} \end{bmatrix} \begin{bmatrix} 1 - C_{11} & -C_{12} \\ -C_{21} & 1 - C_{22} \end{bmatrix}^{-1} \begin{bmatrix} t_{1f} \\ t_{2f} \end{bmatrix} \tag{9-1}$$

$$C_{jk} = \sum_{l=1,2} a_{jl} a_{lk} e^{i(\phi_l + \phi_k)} \tag{9-2}$$

where $a_{mn}$ represents all reflection/swapping coefficients, and $\phi_j = (2\pi L/\lambda_0) n_{eff,j}$ is the phase accumulated during one-way propagation through a MRIM slab with length $L$ in the $j^{th}$ channel. For a special case where two types of optical channels are identical, Eq. 9 is equivalent to the well-known transmission of a conventional Fabry-Perot etalon: $E_t/E_0 = (t_{fx} t_{xf} e^{i\phi_x})/(1 - r_{xx}^2 e^{2i\phi_x})$, where $t_{xf}$, $t_{fx}$ and $r_{xx}$ are the standard Fresnel transmission and reflection coefficients at the interface between free space and material $x$. Using Eq. 9, in Fig. 4(b) we plot the transmission of our two-channel MRIM Fabry-Perot etalon as a function of the cavity length $L$. Comparing to two transparent dielectric etalons with refractive indices matching the effective indices of the MRIM, the MRIM etalon yields an enhanced density of resonant modes, which are close to the transmission resonances of the two dielectric etalons Fig. 4(b vs. d). We note that due to the coupling of two propagation modes via swapping, the MRIM etalon cannot be treated as a "superimposed etalon" by adding up the fields of two dielectric etalons (red curve in Fig. 4(d)), as we did for the "superimposed lens" in Fig 3(c).

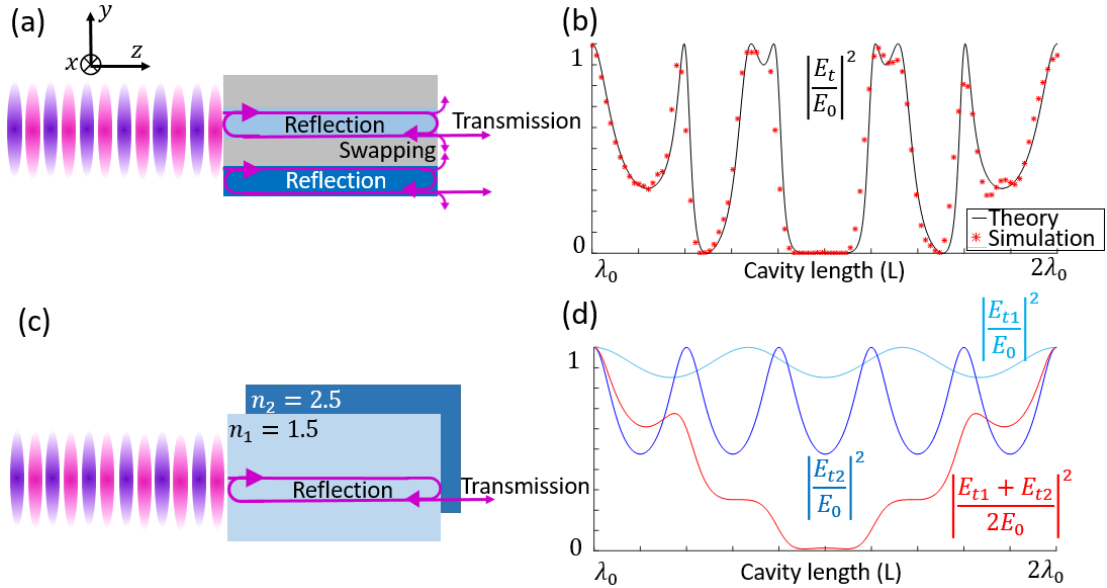

Fig 4. (a): Normal-incidence p-polarized light (with electric field $E_0$ along the y axis) is transmitted through a Fabry-Perot-like etalon comprising a MRIM with two optical channels, and generates transmitted field $E_t$ as shown in (b). (c): The same incident light is independently transmitted through two Fabry-Perot etalons comprising homogeneous materials, generating transmitted fields $E_{t1}$ and $E_{t2}$, respectively, as shown in (d). Even though the MRIM in (a) and the homogeneous materials in (c) have same set of refractive indices, the transmitted field in (b), is different from the sum of the transmitted fields in (d) (red line). The theoretical result in (b) is calculated using Eq. 9, where the MRIM contains dielectric layers with $n_1 = 1.5$ and $d_1 = 100\ nm$, and $n_2 = 2.5$ and $d_2 = 150\ nm$, separated by PEC layers of $100\ nm$ thickness. The simulation result in (b) is obtained using FDTD for an incident plane wave with free-space wavelength $\lambda_0 = 8\ \mu m$ and periodic boundary conditions along the y axis.

In conclusion, we demonstrate a new degree of freedom for the refractive index of metamaterials, in addition to dispersion and birefringence, using subwavelength waveguide arrays. Refractive components made of this kind of multi-refractive-index metamaterial (MRIM) can behave similarly to multiple superimposed conventional refractive components of the same geometry but comprising different refractive indices. We simulated several common optical components, substituting a MRIM for conventional transparent dielectrics, including a prism, a cylindrical lens, and a Fabry-Perot etalon. Given a linearly polarized, single-wavelength incident beam, the prism yields multiple distinct refracted beams, the lens multiple focal points, and the Fabry-Perot etalon an enhanced density of resonant modes. Our result can be generalized to metamaterials comprising a variety of deep-subwavelength waveguides with either one- or two-dimensional confinement, as long as the modes have no cutoff frequency (e.g., coaxial waveguides [15]). The ability to design optical components out of materials with multiple simultaneous refractive index values may enable new functionalities, such as synchronous imaging of multiple focal planes.

M. K. acknowledges financial support from the Wisconsin Alumni Research Foundation through the Fall Competition and from the Air Force Office of Scientific Research (FA9550-18-1-0146). B. G. acknowledges support from the National Science Foundation through a Graduate Research Fellowship. Z.Y. is grateful for helpful discussion with Ming Zhou.

# Supplementary material:

## Multi-refractive-index metamaterials based on subwavelength waveguide arrays

Zhaoning Yu[1,3], Chenghao Wan[2], Jad Salman[3], Bradley S. Gundlach[3], Yuzhe Xiao[3], Zongfu Yu[3], Mikhail A. Kats[3,2,1*]

[1]Department of Physics, [2]Department of Materials Science and Engineering, [3]Department of Electrical and Computer Engineering, University of Wisconsin-Madison, Madison, Wisconsin, USA 53706

*Email address: mkats@wisc.edu

## 1) Light with wavelength $\lambda_0 = 16\ \mu m$ refracted by an MRIM/air interface

Fig. 1(b) in the main text shows linearly polarized monochromatic light refracted by a prism/air interface, where the prism comprises a multi-refractive-index metamaterial (MRIM). Using the finite-difference time-domain (FDTD) method, Fig. 1(c) shows the simulated far-field angular distribution of the refracted light with wavelength $\lambda_0 = 8\ \mu m$, whose second peak $\theta'_{t,2} = 64°$ is 8% away from the ray-optics-like prediction $\theta_{t,2} = 59°$.

As a comparison, in Fig. S1(c) we show the simulated refracted far field given a longer wavelength $\lambda_0' = 16\ \mu m$. The far-field angular distribution has the second peak of $\theta'_{t,2} = 60°$, which is $< 2\%$ away from $\theta_{t,2}$.

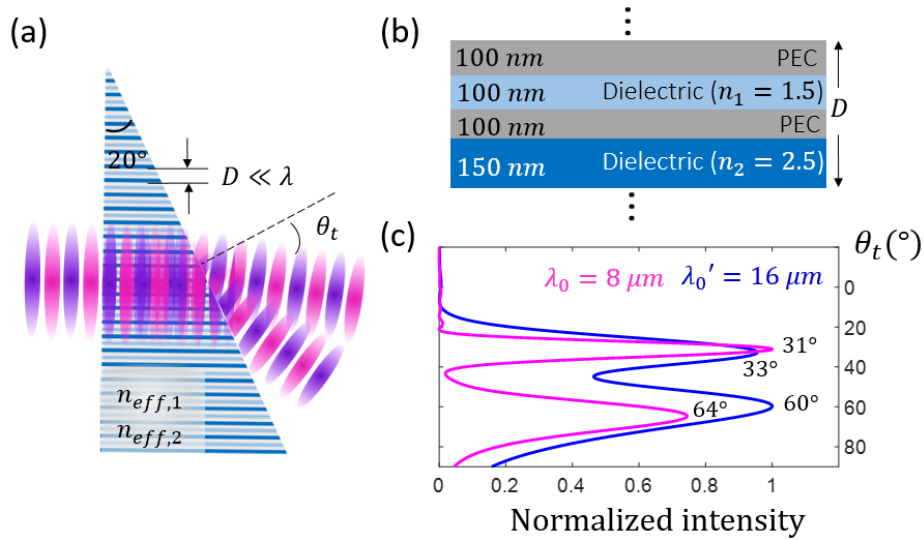

Fig. S1 (a) Linearly polarized monochromatic light refracted by a MRIM/air interface; this figure is the same as Fig. 1(b) in the main text. (b) One period of the MRIM comprising perfect dielectric conductor (PEC) walls and two lossless dielectrics; this figure is the same as Fig. 2(a) in the main text. (c) The simulated far-field angular distribution of the refracted light using incident light with different free-space wavelengths: $\lambda_0 = 8\ \mu m$ and $\lambda_0' = 16\ \mu m$. (c) The far-field distributions, simulated using FDTD, where the incident light is a p-polarized Gaussian beam at normal incidence.

## 2) Light of oblique incidence focused by a cylindrical MRIM lens

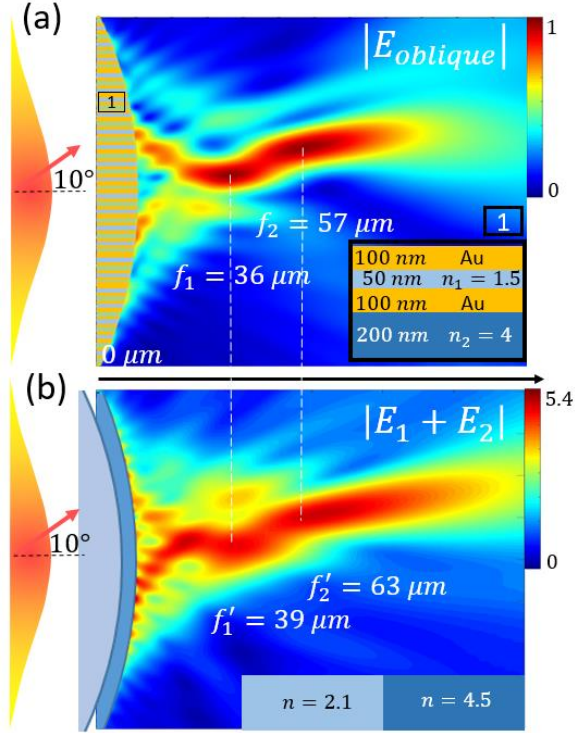

Fig S2. FDTD simulations of incidence light focused by a cylindrical lens. (a): Two foci are generated by our lens comprising a MRIM with thick metal walls, shown in the inset of (a). This panel is the same as Fig. 3(c) in the main text. (b): A field distribution similar to that of (a) is found by coherently adding the fields from two lenses of the same size and shape as in (a), comprising of two different homogeneous transparent dielectrics with refractive indices corresponding to the two effective indices of the MRIM. All lenses are of same shape with radius of curvature $R = 85\ \mu m$ and width $W = 80\ \mu m$. The gold (Au) refractive index is taken to be $n_{Au} = 8.5 + 46.4i$. The incident light is a p-polarized Gaussian beam with $\lambda_0 = 8\ \mu m$ and beam waist $w_0 = 35\ \mu m$.

## 3) Shift of the focal length in small lenses

For $n_{eff,1} = 2.1$ and $n_{eff,2} = 4.5$ and a radius of curvature $R = 85\ \mu m$, the focal lengths given by the Lensmaker's formula (Eq. 2) are:

$$\begin{bmatrix} f_1^{cal} \\ f_2^{cal} \end{bmatrix} = \frac{R}{\begin{bmatrix} n_{eff,1} \\ n_{eff,2} \end{bmatrix} - 1} = \begin{bmatrix} 77 \\ 24 \end{bmatrix}\ \mu m \qquad (S\text{-}1),$$

The Fresnel numbers of the lens are:

$$\begin{bmatrix} N_1 \\ N_2 \end{bmatrix} = \frac{a^2}{\lambda_0 \begin{bmatrix} f_1^{cal} \\ f_2^{cal} \end{bmatrix}} = \begin{bmatrix} 2.6 \\ 8.3 \end{bmatrix} \quad \text{(S-2)},$$

where the radius of the aperture is $a = W/2 = 40\ \mu m$, and $\lambda_0 = 8\ \mu m$.

To calculate the shift of the focal length due to diffraction, where the focal point is defined as the field maximum along the optic axis, we use Eq. 2.18 of ref. [S1]:

$$\frac{\Delta f}{f^{cal}} = \frac{u_N}{2\pi N - u_N} \quad \text{(S-3)},$$

where $u_N$ is the solution of a parametric equation (Eq. 3.11 of ref. [S1]) in the range of $-2\pi$ to $0$:

$$\frac{\tan\frac{u_N}{4}}{\frac{u_N}{4}} = 1 - \frac{u_N}{2\pi N} \quad \text{(S-4)},$$

where the Fresnel number $N$ is either $N_1$ or $N_2$ as calculated in Eq. S-2. We find the numerical solutions of Eq. S-4 to be $u_{N1} = -2.5, u_{N2} = -0.9$. Using calculated $u_N$ and $N$, Eq. S-3 gives the focal length shift: $\Delta f_1/f_1^{cal} = -13\%$ and $\Delta f_2/f_2^{cal} = -2\%$, respectively. The negative sign indicates a decrease in both focal lengths; *i.e.*, the foci are shifted closer to the lens.

**4) The transmission, reflection and swapping coefficients at the interface between free space and a MRIM**

i) Light from free space entering a two-index MRIM made with PEC walls

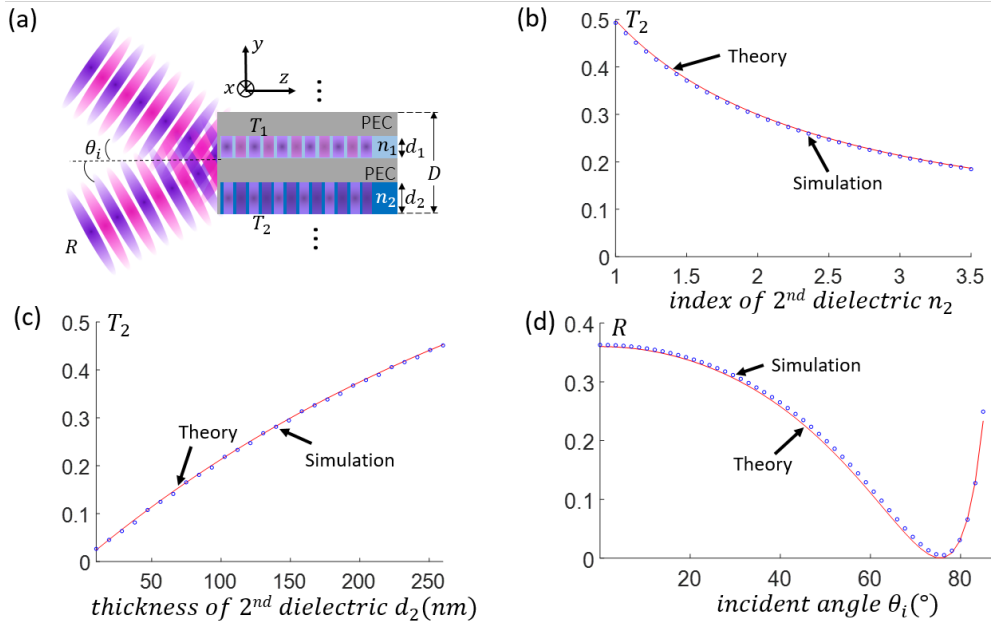

Fig. S3. (a) P-polarized light from free space with incident angle $\theta_i$ is reflected and transmitted to different optical channel of a MRIM with PEC walls. (b), (c), and (d): the transmittance and reflectance of the structure in (a) as a function of the index ($n_2$) and thickness ($d_2$) of the $2^{nd}$ dielectric layer, and of the incident angle ($\theta_i$), respectively. All MRIMs in (b), (c), and (d) have the same period $D = 400\ nm$, $n_1 = 1.5$, and thickness $d_1 = 100\ nm$. MRIMs in (b) have $d_2 = 100\ nm$, MRIMs in (c) have $n_2 = 3$, both with normal incidence. MRIMs in (d) have $d_2 = 100\ nm$ and $n_2 = 3$. Theoretical results (red line) were calculated using Eq. 5 and $T_2 = n_2 t_{f2}^2 d_2/D$ (Eq. 4.57 of ref. [S2], modified using cross-section area ratio $d_2/D$). FDTD simulation results (blue circles) were obtained with an incident plane wave of $\lambda_0 = 8\ \mu m$ and Bloch boundary conditions along the y axis.

### ii) Light exiting a two-index MRIM made with PEC walls

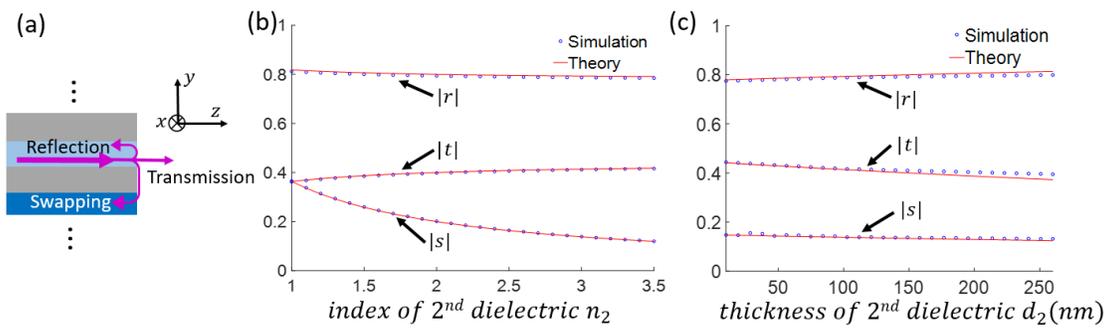

Fig. S4. (a) Light in the TEM mode in one optical channel of a MRIM is reflected, transmitted to free space, and swapped over the other optical channel. (b) and (c): the reflection ($r$), transmission ($t$), and swapping coefficient ($s$) as a function of the index ($n_2$) and thickness ($d_2$) of the second dielectric of the MRIM. All MRIMs in (b) and (c) have the same period $D = 400\ nm$, $n_1 = 1.5$, and thickness $d_1 = 100\ nm$. MRIMs in (b) have $d_2 = 100\ nm$; MRIMs in (c) have $n_2 = 3$, both with normal incidence. Theoretical results (red lines) were calculated using Eq. 8, and the FDTD simulation results (blue circles) were obtained using a subwavelength mode source across the first optical channel with $\lambda_0 = 8\ \mu m$, and periodic boundary conditions along the $y$ axis.

## 5) Transmittance and reflectance of multiple-channel Fabry-Perot etalons

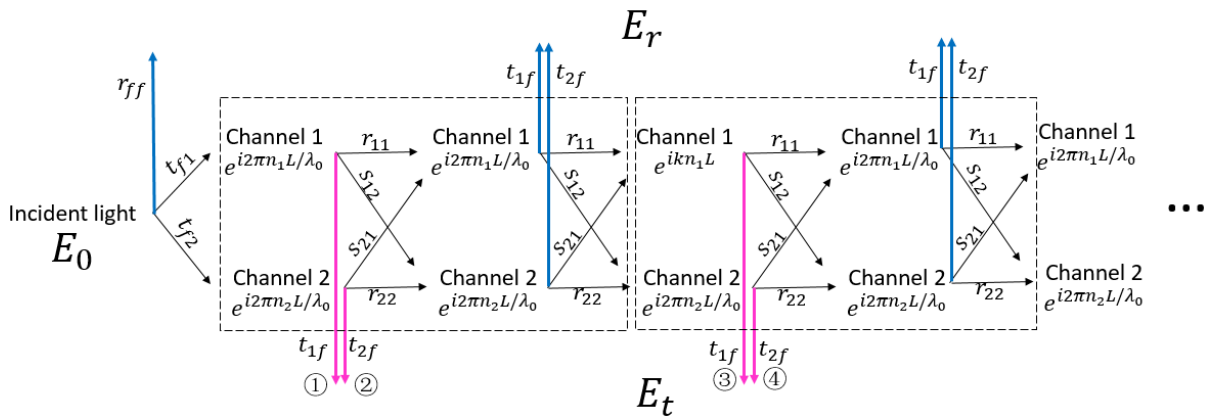

Fig. S5. Schematic of the possible optical paths when light from free space with wavelength $\lambda_0$ is incident on a Fabry-Perot-like etalon made with a MRIM that has 2 optical channels corresponding to propagation mode indices $n_1$, $n_2$. Fields contributing to the total reflection are shown as going up, while those contributing to the total transmission are shown as going to the bottom. Each doted box groups two sequential optical channels and represents a round-trip in the etalon. The Fabry-Perot-like etalon has the cavity length $L$.

A multi-channel Fabry-Perot-like etalon comprising a MRIM slab is shown in Fig. 4(a) in the main text. Due to swapping between channels, a multi-channel etalon cannot be treated as a "superimposed etalon" by manually adding up the field of two dielectric etalons (red curve in Fig. 4(d) in the main text). Instead, it has more possible optical paths, which are shown in Fig. S5. The transmitted field, for example, contains light that has travelled through 1, 3, 5, ... channels, and can be expressed as the sum of the fields from all possible paths (①,②,③,④,... as shown in Fig. S5):

$$\frac{E_t}{E_0} = E_① + E_② + E_③ + E_④ + \cdots = t_{f1}e^{i2\pi n_1 L/\lambda_0} t_{1f} + t_{f2}e^{i2\pi n_2 L/\lambda_0} t_{2f} +$$
$$[t_{f1}e^{i2\pi n_1 L/\lambda_0}\left(r_{11}e^{i2\pi n_1 L/\lambda_0}r_{11}e^{i2\pi n_1 L/\lambda_0} + s_{12}e^{i2\pi n_2 L/\lambda_0}s_{21}e^{i2\pi n_1 L/\lambda_0}\right)t_{1f} +$$
$$t_{f2}e^{i2\pi n_2 L/\lambda_0}\left(s_{21}e^{i2\pi n_1 L/\lambda_0}r_{11}e^{i2\pi n_1 L/\lambda_0} + r_{22}e^{i2\pi n_2 L/\lambda_0}s_{21}e^{i2\pi n_1 L/\lambda_0}\right)t_{1f}] +$$
$$[t_{f1}e^{i2\pi n_1 L/\lambda_0}\left(r_{11}e^{i2\pi n_1 L/\lambda_0}s_{12}e^{i2\pi n_2 L/\lambda_0} + s_{12}e^{i2\pi n_2 L/\lambda_0}r_{22}e^{i2\pi n_2 L/\lambda_0}\right)t_{2f} +$$
$$t_{f2}e^{i2\pi n_2 L/\lambda_0}\left(s_{21}e^{i2\pi n_1 L/\lambda_0}s_{12}e^{i2\pi n_2 L/\lambda_0} + r_{22}e^{i2\pi n_2 L/\lambda_0}r_{22}e^{i2\pi n_2 L/\lambda_0}\right)t_{2f}] + \cdots$$

(S-5)

We categorize an optical path into one of four categories by its initial and final optical channel: a) light initially enters the first optical channel and exits to free space from the first optical channel, b) light initially enters the first optical channel but exits to free space from the second optical channel, c) light initially enters the second optical channel but exits to free space from the first optical channel, and d) light enters and exists to free space from the second optical channel. We can rewrite Eq. S-5 as:

$$\frac{E_t}{E_0} = t_{f1}e^{i2\pi n_1 L/\lambda_0} t_{1f}\left[1 + \left(r_{11}e^{i2\pi n_1 L/\lambda_0}r_{11}e^{i2\pi n_1 L/\lambda_0} + s_{12}e^{i2\pi n_2 L/\lambda_0}s_{21}e^{i2\pi n_1 L/\lambda_0}\right) + \cdots\right] +$$
$$t_{f1}e^{i2\pi n_1 L/\lambda_0} t_{2f}\left[\left(r_{11}e^{i2\pi n_1 L/\lambda_0}s_{12}e^{i2\pi n_2 L/\lambda_0} + s_{12}e^{i2\pi n_2 L/\lambda_0}r_{22}e^{i2\pi n_2 L/\lambda_0}\right) + \cdots\right] +$$
$$t_{f2}e^{i2\pi n_2 L/\lambda_0} t_{1f}\left[\left(s_{21}e^{i2\pi n_1 L/\lambda_0}r_{11}e^{i2\pi n_1 L/\lambda_0} + r_{22}e^{i2\pi n_2 L/\lambda_0}s_{21}e^{i2\pi n_1 L/\lambda_0}\right) + \cdots\right] +$$
$$t_{f2}e^{i2\pi n_2 L/\lambda_0} t_{2f}\left[1 + \left(s_{21}e^{i2\pi n_1 L/\lambda_0}s_{12}e^{i2\pi n_2 L/\lambda_0} + r_{22}e^{i2\pi n_2 L/\lambda_0}r_{22}e^{i2\pi n_2 L/\lambda_0}\right) + \cdots\right]$$

(S-6)

In Eq. S-6, only light that has travelled through one channel ((①,② in Fig. S5) and three channels (③,④ in Fig. S5) are written out, higher order terms are included in the ellipsis. To account for higher order terms which represent 5, 7, 9, ... channels, we define recursive coefficients:

$$C_{11} \triangleq r_{11}e^{i2\pi n_1 L/\lambda_0}r_{11}e^{i2\pi n_1 L/\lambda_0} + s_{12}e^{i2\pi n_2 L/\lambda_0}s_{21}e^{i2\pi n_1 L/\lambda_0},$$
$$C_{12} \triangleq r_{11}e^{i2\pi n_1 L/\lambda_0}s_{12}e^{i2\pi n_2 L/\lambda_0} + s_{12}e^{i2\pi n_2 L/\lambda_0}r_{22}e^{i2\pi n_2 L/\lambda_0},$$
$$C_{21} \triangleq s_{21}e^{i2\pi n_1 L/\lambda_0}r_{11}e^{i2\pi n_1 L/\lambda_0} + r_{22}e^{i2\pi n_2 L/\lambda_0}s_{21}e^{i2\pi n_1 L/\lambda_0},$$
$$C_{22} \triangleq s_{21}e^{i2\pi n_1 L/\lambda_0}s_{12}e^{i2\pi n_2 L/\lambda_0} + r_{22}e^{i2\pi n_2 L/\lambda_0}r_{22}e^{i2\pi n_2 L/\lambda_0}$$

(S-7)

The recursive coefficient $C_{jk}$ is the electric field ratio ($E_{final}/E_{initial}$) when adding two more optical channels, starting from the $j^{th}$ channel and ending with the $k^{th}$ channel.

We can simplify Eq. S-7 if we denote all $r_{jk}$, $s_{jk}$ as $a_{jk}$, keeping the subscripts the same. Then all four equations in Eq. S-7 can be written as:

$$C_{jk} = \sum_{l=1,2} a_{jl} a_{lk} e^{i(\phi_l + \phi_k)} \tag{S-8}$$

where $\phi_j = 2\pi L/\lambda_0 * n_{eff,j}$ is the phase accumulated during one-way propagation with cavity length $L$ in the $j^{th}$ channel.

When increasing total number of channels by four starting from the $j^{th}$ channel and ending with the $k^{th}$ channel, there are two possibilities: a) the first two channels also end with the $k^{th}$ channel, and the later two channels both start from and end with the $k^{th}$, b) the first two channels end with the $l^{th}$ ($l \neq k$) channel, and the later two channels start from the $l^{th}$ channel and end with the $k^{th}$ channel. Thus the electric field ratio by adding four more optical channels $D_{jk}$ is related to $C_{jk}$ in the following way:

$$D_{11} = C_{11}C_{11} + C_{12}C_{21},$$
$$D_{12} = C_{12}C_{22} + C_{11}C_{12},$$
$$D_{21} = C_{21}C_{11} + C_{22}C_{21},$$
$$D_{22} = C_{22}C_{22} + C_{21}C_{12}$$

$$\tag{S-9}$$

Eq. S-9 can be rewritten in matrix format:

$$\begin{bmatrix} D_{11} & D_{12} \\ D_{21} & D_{22} \end{bmatrix} = \begin{bmatrix} C_{11} & C_{12} \\ C_{21} & C_{22} \end{bmatrix}^2 \tag{S-10}$$

Based on similar reasoning the electric field ratio by adding six more optical channels $F_{jk}$ is:

$$\begin{bmatrix} F_{11} & F_{12} \\ F_{21} & F_{22} \end{bmatrix} = \begin{bmatrix} D_{11} & D_{12} \\ D_{21} & D_{22} \end{bmatrix} \begin{bmatrix} C_{11} & C_{12} \\ C_{21} & C_{22} \end{bmatrix} = \begin{bmatrix} C_{11} & C_{12} \\ C_{21} & C_{22} \end{bmatrix}^3 \tag{S-11}$$

Using $C_{jk}$, $D_{jk}$, $F_{jk}$, etc., the total transmitted electric field (Eq. S-6), which contains light that has travelled through 1, 1+2, 1+4, 1+6, … number of channels, can be written as:

$$\frac{E_t}{E_0} = t_{f1} e^{ikn_1 L} t_{1f}(1 + C_{11} + D_{11} + F_{11} + \cdots) + t_{f1} e^{ikn_1 L} t_{2f}(C_{12} + D_{12} + F_{12} + \cdots) +$$
$$t_{f2} e^{ikn_2 L} t_{1f}(C_{21} + D_{21} + F_{21} + \cdots) + t_{f2} e^{ikn_2 L} t_{2f}(1 + C_{22} + D_{22} + F_{22} + \cdots)$$

$$\tag{S-12}$$

Eq. S-12 can be rewritten in matrix format:

$$\frac{E_t}{E_0} = \begin{bmatrix} t_{f1} e^{i\phi_1} & t_{f2} e^{i\phi_2} \end{bmatrix} \left( \begin{bmatrix} 1 & 0 \\ 0 & 1 \end{bmatrix} + \begin{bmatrix} C_{11} & C_{12} \\ C_{21} & C_{22} \end{bmatrix} + \begin{bmatrix} D_{11} & D_{12} \\ D_{21} & D_{22} \end{bmatrix} + \begin{bmatrix} F_{11} & F_{12} \\ F_{21} & F_{22} \end{bmatrix} + \cdots \right) \begin{bmatrix} t_{1f} \\ t_{2f} \end{bmatrix}$$

$$\tag{S-13}$$

We note that:

$$\begin{bmatrix} 1 & 0 \\ 0 & 1 \end{bmatrix} + \begin{bmatrix} C_{11} & C_{12} \\ C_{21} & C_{22} \end{bmatrix} + \begin{bmatrix} D_{11} & D_{12} \\ D_{21} & D_{22} \end{bmatrix} + \begin{bmatrix} F_{11} & F_{12} \\ F_{21} & F_{22} \end{bmatrix} + \cdots = \sum_{m=0}^{\infty} \begin{bmatrix} C_{11} & C_{12} \\ C_{21} & C_{22} \end{bmatrix}^m$$

$$= \begin{bmatrix} 1 - C_{11} & -C_{12} \\ -C_{21} & 1 - C_{22} \end{bmatrix}^{-1} \tag{S-14}$$

Thus Eq. S-12 can be written as:

$$\frac{E_t}{E_0} = \begin{bmatrix} t_{f1}e^{i\phi_1} & t_{f2}e^{i\phi_2} \end{bmatrix} \begin{bmatrix} 1 - C_{11} & -C_{12} \\ -C_{21} & 1 - C_{22} \end{bmatrix}^{-1} \begin{bmatrix} t_{1f} \\ t_{2f} \end{bmatrix} \tag{S-15}$$

The transmittance of two-channel Fabry-Perot etalons calculated from Eq. S-15 agrees with simulations, as shown in Fig. 4(b) in the main text when plotting as a function of cavity length $L$. We note that due to the coupling between two propagation modes in the two-channel etalons, the resonant peaks are slightly shifted from those of etalons comprising homogeneous materials (blue and purple curve in Fig. 4(d)).

For the case of a conventional Fabry-Perot etalon made with single-index dielectric (denoted as structure x), the two optical channels in Eq. S-15 are identical, which means: a) the swapping vanishes, *i.e.*, $s_{12} = s_{21} = 0$, b) the reflection/transmission coefficients are identical for 2 channels, *i.e.*, $r_{xx} = r_{11} = r_{22}$, $t_{fx} = t_{f1} = t_{f2}$, and $t_{xf} = t_{1f} + t_{2f} = 2t_{1f} = 2t_{2f}$. Then recursive coefficients (Eq. S-7) become: $C_{11} = C_{22} = r_{xx}^2 e^{2i\phi_x}$, $C_{12} = C_{21} = 0$. As a result, Eq. S-14 modified for single-channel Fabry-Perot etalon is:

$$\frac{E_t}{E_0} = \begin{bmatrix} t_{fx}e^{i\phi_x} & t_{fx}e^{i\phi_x} \end{bmatrix} \begin{bmatrix} 1 - r_{xx}^2 e^{2i\phi_x} & 0 \\ 0 & 1 - r_{xx}^2 e^{2i\phi_x} \end{bmatrix}^{-1} \begin{bmatrix} \frac{t_{xf}}{2} \\ \frac{t_{xf}}{2} \end{bmatrix} = \frac{t_{fx}t_{xf}e^{i\phi_x}}{1 - r_{xx}^2 e^{2i\phi_x}} \tag{S-16}$$

Eq. S-16 is indeed the transmission coefficient of conventional Fabry-Perot etalon, which can be seen if it is written as $E_t/E_0 = t_{fx}t_{xf}(e^{i\phi_x} + r_{xx}^2 e^{3i\phi_x} + r_{xx}^4 e^{5i\phi_x} + \cdots)$, containing light that travels 1, 3, 5… cavity lengths.

The reflected field can be calculated in a similar fashion with the transmitted field, and we provide here the final result:

$$\frac{E_r}{E_0} = r_{ff} + \begin{bmatrix} t_{f1}e^{i\phi_1} & t_{f2}e^{i\phi_2} \end{bmatrix} \begin{bmatrix} r_{11} & s_{12} \\ s_{21} & r_{22} \end{bmatrix} \begin{bmatrix} 1 - C_{11} & -C_{12} \\ -C_{21} & 1 - C_{22} \end{bmatrix}^{-1} \begin{bmatrix} t_{1f}e^{i\phi_1} \\ t_{2f}e^{i\phi_2} \end{bmatrix} \tag{S-16}$$

The transmitted/reflected field for Fabry-Perot-like etalons that have more than two optical channels can be calculated with the same approach as for two-channel etalons, with $m \times m$ recursion matrix ($C$) with element:

$$C_{jk} = \sum_{l=1,2,\ldots,m} a_{jl}a_{lk}e^{i(\phi_l+\phi_k)} \tag{S-17}$$

The corresponding transmitted/reflected field becomes:

$$\frac{E_t}{E_0} = \begin{bmatrix} t_{f1}e^{i\phi_1} & t_{f1}e^{i\phi_1} & \cdots & t_{fm}e^{i\phi_m} \end{bmatrix} (I-C)^{-1} \begin{bmatrix} t_{1f} \\ t_{2f} \\ \vdots \\ t_{mf} \end{bmatrix}$$

$$\frac{E_r}{E_0} = r_{ff} + \begin{bmatrix} t_{f1}e^{i\phi_1} & t_{f1}e^{i\phi_1} & \cdots & t_{fm}e^{i\phi_m} \end{bmatrix} \begin{bmatrix} r_{11} & s_{12} & \cdots & s_{1m} \\ s_{21} & r_{22} & \cdots & s_{2m} \\ \vdots & \vdots & \ddots & \vdots \\ s_{m1} & r_{m2} & \cdots & r_{mm} \end{bmatrix} (I-C)^{-1} \begin{bmatrix} t_{1f}e^{i\phi_1} \\ t_{2f}e^{i\phi_2} \\ \vdots \\ t_{mf}e^{i\phi_m} \end{bmatrix}$$

(S-18)

where I is an $m \times m$ identity matrix.

## 6) Selecting dielectric layers for a MRIM lens to obtain separable foci of comparable field magnitude

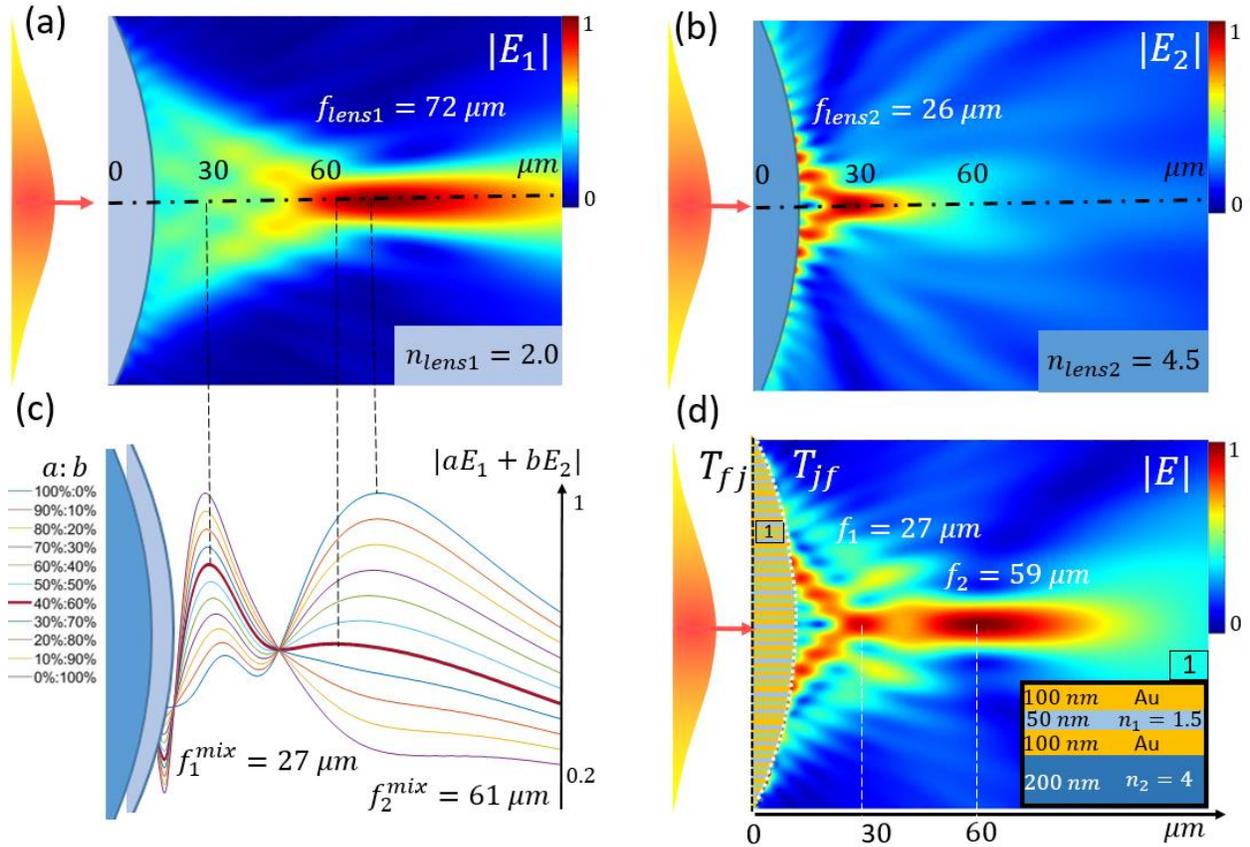

Fig S6. FDTD simulations of light focused by a cylindrical lens. (a) & (b): One focus is generated by homogeneous transparent dielectrics with refractive index $n_{lens1} = 2.0$ and $n_{lens2} = 4.5$, respectively. (d): Two foci are generated by our lens comprising a MRIM with thick metal walls, shown in the inset of (d). Foci at very similar positions to (d) are found at (c) the coherently summed field along the optical axis of (a) and (b) with ratio $40\%$: $60\%$, which is approximately the ratio achieved by the MRIM lens in (d). All lenses are of the same plano-convex shape with radius of curvature $R = 85\ \mu m$, width $W = 80\ \mu m$, and minimum thickness $t_e = 1\ \mu m$ at the edge. The refractive index of gold is taken to be $n_{Au} = 8.5 + 46.4i$ [S2]. The incident light is a p-polarized Gaussian beam with $\lambda_0 = 8\ \mu m$ and beam waist $w_0 = 35\ \mu m$.

We start with two dielectric cylindrical lenses that have different refractive indices; e.g., $n_{lens1} = 2.0$ and $n_{lens2} = 4.5$ in Fig. S6(a, b). Both lenses have the same plano-convex shape, which will be the shape of the potential MRIM lens, with radius of curvature $R = 85\ \mu m$, width $W = 80\ \mu m$, and minimum thickness $t_e = 1\ \mu m$ at the edge.

Using the FDTD method, we simulate light focused by the dielectric lenses separately (the normalized electric field magnitude is shown in Fig. S6(a, b)), with incident light being a p-polarized Gaussian beam with $\lambda_0 = 8\ \mu m$ and beam waist $w_0 = 35\ \mu m$. We then extract the normalized field along the optic axis and add them coherently with different ratios (Fig. S6(c)), with 100%:0 and 0:100% being the electric field of lens1 ($f_{lens1} = 72\ \mu m$) and lens2 ($f_{lens2} = 26\ \mu m$), respectively. As shown in Fig. S6(c), separable foci with comparable field magnitude are found when mixing with 40% or more of the field of lens1 ($a \geq 40\%$). As an example, we chose the mixing ratio of the normalized field to be 40%: 60%, where the mixed field has foci at $f_1^{mix} = 27\ \mu m$ and $f_2^{mix} = 61\ \mu m$ (bolded curve in Fig. S6(c)).

The field distribution in Fig. S6(c) can be reproduced by using an ideal MRIM lens with effective indices $n_{eff,1}^{ideal} = 2.0$ and $n_{eff,2}^{ideal} = 4.5$, and with fractions of power to each refractive index being $P_1/P_2 = (40\%/60\%)^2$. The desired effective indices can be achieved by choosing dielectric layers with proper refractive index. The fractions of power to each index can be adjusted by tuning thickness combination of the two dielectric layers, which changes the transmission coefficient to each optical channel (Eq. 5 and 8 for MRIMs based on PEC).

For example, the MRIM lens in Fig. 3(a, c) contains gold walls with refractive index $n_{Au} = 8.5 + 46.4i$ [3] and thickness $d_{Au} = 100\ nm$. To achieve similar effective indices to those described above, we select dielectric layers of lower index based on Fig. 2(b); e.g., $n_1 = 1.5$ and $n_2 = 4.0$. Then we select the thickness of dielectric layers to achieve desired fractions of power to each index. To do so we made two approximations. First, we approximate the total transmitted power as the product of the transmittances at the two boundaries of the lens ($T_{fj}$ and $T_{jf}$ in Fig. S6(d)). This approximation ignores secondary bounces inside the lens and the difference of loss between light propagating in the two channels. Second, we approximate each transmittance as that of MRIMs with PEC walls, which can be calculated using the transmission coefficients ($t_{fj}$ and $t_{jf}$) given in Eq. 5 and 8. With these assumptions, the ratio of power to each index becomes:

$$\frac{P_1}{P_2} = \left(\frac{E_1}{E_2}\right)^2 = \left(\frac{E_{inc} t_{f1} t_{1f}}{E_{inc} t_{f2} t_{2f}}\right)^2 = \left(\frac{n_2 d_1}{n_1 d_2}\right)^2 \quad \text{(S-19)},$$

where $E_{inc}$ represents the electric field of incident beam and $d_j$ is the thickness of $j^{th}$ dielectric layer. With $P_1/P_2 = (40\%/60\%)^2$ and $n_1/n_2 = 1.5/4$, the thickness ratio of dielectric layers should be $d_1/d_2 = 1/4$. We choose $d_1 = 50\ nm$ and $d_2 = 200\ nm$ to stay close to the thickness of the gold layers. The resulting MRIM lens in Fig. 3(a, c) with such dielectric layer combination has effective indices of $n_{eff,1} = 2.1, n_{eff,2} = 4.5$ compared to $n_{eff,1}^{ideal} = 2.0, n_{eff,2}^{ideal} = 4.5$, and focal lengths of $f_1 = 27$ μm and $f_2 = 59\ \mu m$, compared to $f_1^{mix} = 27\ \mu m$ and $f_2^{mix} = 61\ \mu m$.

**Supplementary References**